# Enhancement of the Nernst effect by stripe order in a high-$T_c$ superconductor

Olivier Cyr-Choinière [1], R. Daou [1], Francis Laliberté [1], David LeBoeuf [1],

Nicolas Doiron-Leyraud [1], J. Chang [1], J.-Q. Yan [2,†], J.-G. Cheng [2], J.-S. Zhou [2],

J.B. Goodenough [2], S. Pyon [3], T. Takayama [3], H. Takagi [3,4], Y. Tanaka [5,3]

& Louis Taillefer [1,6]

1 Département de physique and RQMP, Université de Sherbrooke, Sherbrooke, Québec
J1K 2R1, Canada

2 Texas Materials Institute, University of Texas at Austin, Austin, Texas 78712, USA

3 Department of Advanced Materials, University of Tokyo, Kashiwa 277-8561, Japan

4 RIKEN (The Institute of Physical and Chemical Research), Wako, 351-0198, Japan

5 RIKEN SPring8 Center, Hyogo 679-5148, Japan

6 Canadian Institute for Advanced Research, Toronto, Ontario M5G 1Z8, Canada

**The Nernst effect in metals is highly sensitive to two kinds of phase transition: superconductivity and density-wave order[1]. The large positive Nernst signal observed in hole-doped high-$T_c$ superconductors[2] above their transition temperature $T_c$ has so far been attributed to fluctuating superconductivity[3]. Here we show that in some of these materials the large Nernst signal is in fact caused by stripe order, a form of spin / charge modulation[4] which causes a reconstruction of the Fermi surface[5]. In LSCO doped with Nd or Eu, the onset of stripe order causes the Nernst signal to go from small and negative to large and positive, as revealed either by**

---

† Present address : Ames Laboratory, Ames, Iowa 50011 USA

**lowering the hole concentration across the quantum critical point in Nd-LSCO (refs. 6, 7, 8), or lowering the temperature across the ordering temperature in Eu-LSCO (refs. 9, 10). In the latter case, two separate peaks are resolved, respectively associated with the onset of stripe order at high temperature and superconductivity near $T_c$. This sensitivity to Fermi-surface reconstruction makes the Nernst effect a promising probe of broken symmetry in high-$T_c$ superconductors.**

The Nernst effect is the development of a transverse electric field $E_y$ across the width ($y$-axis) of a metallic sample when a temperature gradient $\partial T / \partial x$ is applied along its length ($x$-axis) in the presence of a transverse magnetic field $B$ (along the $z$-axis). Two mechanisms can give rise to a Nernst signal, defined as $N = E_y / ( \partial T / \partial x )$ ( ref. 1). The first is superconducting fluctuations, of either phase or amplitude[1,3], which can only be positive[1]; the second is due to mobile charge carriers, given by[1]:

$$N = - e \, L_0 \, T \; \partial( \, \sigma_{xy} / \sigma_{xx} \, ) \, / \, \partial\varepsilon \, |_{\, \varepsilon \, = \, \varepsilon F} \quad , \qquad (1)$$

where $e$ is the electron charge, $L_0 \equiv \pi^2 / 3 \, ( \, k_B / e \, )^2$, $T$ is the temperature, $\varepsilon$ is the energy, $\varepsilon_F$ the Fermi energy, $\sigma_{xy}$ is the (transverse) Hall conductivity, and $\sigma_{xx}$ the (longitudinal) electrical conductivity. This quasiparticle Nernst signal can be either positive or negative.

While in a single band metal $N$ is generally small, in a multi-band metal it can be large[1], as indeed it is in semi-metals, where the Nernst coefficient $\nu \equiv N / B$ is typically very large[1,11]. This implies that the quasiparticle Nernst coefficient should generically undergo a pronounced rise when the Fermi surface of a single-band metal is reconstructed into several pieces by the onset of some density-wave-like order. This is indeed what happens in metals like $URu_2Si_2$ (ref. 12) as they enter a semi-metallic ordered state[1,11].

Evidence that the Fermi surface of high-$T_c$ superconductors undergoes some reconstruction in the underdoped regime came recently from the observation of low-frequency quantum oscillations in $YBa_2Cu_3O_y$ (YBCO) (ref. 13), thought to arise from orbits around a small electron-like Fermi pocket[14]. Indeed, the standard mechanism for producing small electron pockets out of a large hole-like Fermi surface is the onset of some density-wave order which breaks translational symmetry[15,5]. Within such a density-wave scenario, the Nernst coefficient of a single-band metal like $La_{2-x}Sr_xCuO_4$ (LSCO) would be expected to undergo a pronounced increase as the material is cooled below its ordering temperature. This is precisely what measurements of the Nernst effect in LSCO have revealed: $\nu$ is small (and negative) at high temperature and becomes large (and positive) at low temperature[2,3]. However, this large rise in $\nu(T)$ with decreasing temperature has instead been attributed to a vortex contribution which grows with the approach of superconductivity[3]. How can we discriminate between these two mechanisms – a change in Fermi surface vs superconducting fluctuations? Here we present two experiments which show that in some high-$T_c$ superconductors the onset of "stripe order" – a form of spin /charge modulation – triggers a large enhancement of the Nernst signal. The material used is LSCO with some of the La replaced by either Nd or Eu, a substitution which stabilizes stripe order[7,9].

In the first experiment, we switch stripe order on and off while keeping the superconductivity constant. This was achieved by measuring two samples of $La_{1.6-x}Nd_{0.4}Sr_xCuO_4$ (Nd-LSCO) with comparable $T_c$ ($\approx 20$ K) but hole concentrations on either side of the critical doping $p*$ where stripe order sets in[6,7,8,9], namely at $p = 0.20$ and $p = 0.24$. The Nernst coefficient $\nu$ of this pair of samples is plotted in Fig. 1 as a function of temperature, along with the in-plane resistivity $\rho$ ( $\rho_{xx}$ ) and Hall coefficient $R_H$ ( ~ $\rho_{xy}$ / $B$ ). In the sample with $p = 0.24$, all coefficients are monotonic and featureless, while in the sample with $p = 0.20$, they all show a pronounced and simultaneous rise.

At $p = 0.24$, the fact that $R_H(T \rightarrow 0) = + 1 / e (1 + p)$ shows that the Fermi surface remains a single large hole cylinder down to the lowest temperature[6]. In this case, $\nu$ is field-independent above $T_c$ (see Supplementary Fig. S1) and remains small and negative down to $T \rightarrow 0$, in agreement with previous data from a non-superconducting LSCO sample with $p = x = 0.26$ (ref. 3). This demonstrates that the onset of superconductivity has, by itself, little impact on $\nu$. In dramatic contrast, at $p = 0.20$, $\nu(T)$ rises rapidly below 40 K to become large and positive, until it vanishes when superconductivity sets in. That the upturn in $\nu(T)$ tracks the upturn in $\rho(T)$ provides a second, independent, evidence that the rise in $\nu(T)$ is not caused by incipient superconductivity.

The parallel rise observed in all three coefficients displayed in Fig. 1 demonstrates that the onset of a large positive Nernst coefficient is due to an enhancement of the quasiparticle contribution rooted in a modification of the Fermi surface[6]. In Fig. 1a, we reproduce the NQR "wipe-out fraction" measured on Nd-LSCO at $x = 0.20$ (ref. 7). The precipitous loss of NQR intensity below 40 K is caused by the onset of stripe order[7] (see also ref. 9). The crucial fact that the upturn in all coefficients matches with its onset strongly suggests that stripe order is the cause of the Fermi-surface reconstruction[5].

In a second experiment, we investigate the more underdoped regime in $La_{1.8-x}Eu_{0.2}Sr_xCuO_4$ (Eu-LSCO). In Fig. 1e, we show X-ray diffraction data on Eu-LSCO at $p = 1/8$. The intensity of scattering at the incommensurate stripe wavevector is seen to vanish at $T_{CO} = 80 \pm 10$ K. In Figs. 1f to 1h, we show transport data taken on one sample with $p = 1/8$. It is clear that the pronounced changes in $\rho(T)$, $R_H(T)$ and $\nu(T)$ all coincide with the onset of stripe order, as in Nd-LSCO at $p = 0.20$. Note that stripe ordering at $p = 1/8$ now causes $R_H(T)$ to drop below $T_{CO}$, as opposed to the rise seen at $p = 0.20$. This evolution in the behaviour of $R_H(T)$ is consistent with calculations[16] based on a theory of the Fermi-surface reconstruction by stripe order[17].

In Supplementary Fig. S2, we define $T_v$, the onset of the upturn in $v(T)$, whose doping dependence is plotted in the inset of Fig. 2. Because of the wide separation between $T_v \approx 140$ K and $T_c \approx 10$ K in Eu-LSCO at $p = 1/8$, we can see that $v(T)$ consists of two separate peaks. The evolution of this two-peak structure with doping is shown in Fig. 2. The low-temperature peak, due to superconducting fluctuations, is suppressed by a magnetic field, while the high-temperature peak, due to quasiparticles, is not (see Supplementary Figs. S3 and S4). A similar situation prevails in the electron-doped cuprate $Pr_{2-x}Ce_xCuO_4$, where the Nernst signal separates clearly into a quasiparticle peak at high temperature and a superconducting peak near $T_c$ (ref. 18). In this case, Fermi-surface reconstruction is attributed to antiferromagnetic order[18,19]. A comparison between Eu-LSCO and LSCO shows that the onset of the positive rise in $v(T)$ occurs at a very similar $T_v$ in both materials (see Supplementary Figs. S3 and S5), suggesting a common mechanism of Fermi-surface reconstruction.

In summary, we have resolved two contributions to the Nernst signal in the hole-doped cuprate LSCO, doped with Nd or Eu: one at low temperature, caused by superconducting fluctuations, the other at high temperature, caused by a change in the Fermi surface. In this case, the change in Fermi surface is clearly caused by the onset of stripe order at $T_{CO}$ (ref. 6). The fact that $v(T)$ starts to rise at $T_v \approx 2 \, T_{CO}$ suggests that stripe fluctuations are sufficient to cause $v(T)$ to increase[5]. It will be interesting to investigate whether the same mechanism is also at play in other hole-doped high-$T_c$ superconductors.

[1] Behnia, K. The Nernst effect and the boundaries of the Fermi liquid picture. *J. Phys.: Condens. Matter* **21,** 113101 (2009).

[2] Xu, Z.A. *et al.* Vortex-like excitations and the onset of superconducting phase fluctuation in underdoped La$_{2-x}$Sr$_x$CuO$_4$. *Nature* **406**, 486-488 (2000).

[3] Wang, Y., Li, P. & Ong, N.P. Nernst effect in high-$T_c$ superconductors. *Phys. Rev. B* **73,** 024510 (2006).

[4] Tranquada, J.M. *et al.* Evidence for stripe correlations of spins and holes in copper oxide superconductors. *Nature* **375**, 561-563 (1995).

[5] Taillefer, L. Fermi-surface reconstruction in high-$T_c$ superconductors. *J. Phys.: Condens. Matter* (2009), in press. Preprint at <http://arXiv.org/abs/0901.2313> (2009).

[6] Daou, R. *et al.* Linear temperature dependence of the resistivity and change in Fermi surface at the pseudogap critical point of a high-$T_c$ superconductor. *Nature Phys.* **5**, 31-34 (2009).

[7] Ichikawa, N. *et al.* Local magnetic order vs superconductivity in a layered cuprate. *Phys. Rev. Lett.* **85**, 1738-1741 (2000).

[8] Daou, R. *et al.* Thermopower across the pseudogap critical point of La$_{1.6-x}$Nd$_{0.4}$Sr$_x$CuO$_4$ : evidence of a quantum critical point in a hole-doped high-$T_c$ superconductor. Preprint at <http://arXiv.org/abs/0810.4280> (2008).

[9] Hunt, A.W. *et al.* Glassy slowing of stripe modulation in (La,Eu,Nd)$_{2-x}$(Sr,Ba)$_x$CuO$_4$ : a $^{63}$Cu and $^{139}$La NQR study down to 350 mK. *Phys. Rev. B* **64**, 134525 (2001).

[10] Fink, J. *et al.* Charge order in La$_{1.8-x}$Eu$_{0.2}$Sr$_x$CuO$_4$ studied by resonant soft X-ray diffraction. Preprint at <http://arXiv.org/abs/0805.4352> (2008).

[11] Behnia, K., Méasson, M.-A. & Kopelevich, Y. Nernst effect in semi-metals: the effective mass and the figure of merit. *Phys. Rev. Lett.* **98**, 076603 (2007).

[12] Bel, R. *et al*. Thermoelectricity of $URu_2Si_2$: giant Nernst effect in the hidden-order state. *Phys. Rev. B.* **70**, 220501 (2004).

[13] Doiron-Leyraud, N. *et al*. Quantum oscillations and the Fermi surface in an underdoped high-$T_c$ superconductor. *Nature* **447**, 565-568 (2007).

[14] LeBoeuf, D. *et al*. Electron pockets in the Fermi surface of hole-doped high-$T_c$ superconductors. *Nature* **450**, 533-536 (2007).

[15] Chakravarty, S. From complexity to simplicity. *Science* **319**, 735-736 (2008).

[16] Lin, J. & Millis, A.J. Theory of low-temperature Hall effect in stripe-ordered cuprates. *Phys. Rev. B* **78**, 115108 (2008).

[17] Millis, A.J. & Norman, M.R. Antiphase stripe order as the origin of electron pockets observed in 1/8-hole-doped cuprates. *Phys. Rev. B* **76**, 220503 (2007).

[18] Li, P. & Greene, R.L. Normal-state Nernst effect in electron-doped $Pr_{2-x}Ce_xCuO_4$: superconducting fluctuations and two-band transport. *Phys. Rev. B* **76**, 174512 (2007).

[19] Hackl, A. & Sachdev, S. Nernst effect in the electron-doped cuprates. Preprint at <http://arXiv.org/abs/0901.2348> (2009).

**Supplementary Information** is linked to the online version of the paper at www.nature.com/nature.

**Acknowledgements** We thank K. Behnia, S. Sachdev and A.-M.S. Tremblay for helpful discussions, and J. Corbin for his assistance with the experiments. JC is supported by a Fellowship from the Swiss National Science Foundation. JSZ and JBG were supported by an NSF grant. HT acknowledges MEXT

Japan for a grant-in-aid for scientific research. LT acknowledges support from the Canadian Institute for Advanced Research and funding from NSERC, FQRNT, and a Canada Research Chair.

**Author Contributions** O. C.-C. and R. D. contributed equally to this work.

**Author Information** Reprints and permissions information is available at www.nature.com/reprints. The authors declare no competing financial interests. Correspondence and requests for materials should be addressed to L.T. (louis.taillefer@physique.usherbrooke.ca).

### Figure 1 | Transport coefficients and stripe order in Nd / Eu-LSCO.

**a)** Charge "stripe" ordering in Nd-LSCO at $p = 0.20$, as measured by the loss of NQR intensity (from ref. 9). At dopings $p = 0.12$ and $p = 0.15$, where both X-ray diffraction and NQR were measured on Nd-LSCO, the lost (or "wipe-out") fraction of the intensity present at 100 K tracks the increase in the intensity of superlattice peaks detected with X-rays[9]. At $p = 0.20$, the onset of charge order is $T_{CO} = 40 \pm 6$ K (ref. 9). *Lower left panels*: transport coefficients in two samples of Nd-LSCO, respectively with $p = 0.20$ (red) and at $p = 0.24$ (blue): **b)** in-plane electrical resistivity $\rho$ in a magnetic field $B = 0$ (open symbols) and 15 T (closed symbols) (from ref. 6); **c)** Hall coefficient $R_H$ in 15 T (from ref. 6); **d)** Nernst coefficient $v$ in 10 T (this work). In both samples, $T_c \approx 20$ K in zero field (see panel (b)). Note how at $p = 0.20$ all coefficients show a pronounced and simultaneous upturn starting at a temperature which coincides with the onset of charge order – strong evidence for a scenario of Fermi-surface reconstruction by stripe order as the cause of the large positive Nernst signal. By contrast, at $p = 0.24$, $v(T)$ remains small and negative, unaffected by the onset of superconductivity. **e)** Charge "stripe" ordering in Eu-LSCO at $p = 0.125$ measured by hard (closed symbols; this work) and soft (open symbols; ref. 10)

X-ray diffraction. Error bars on closed symbols represent the standard error on the height of the Gaussian in a Gaussian + background fit to the momentum scan at each temperature. Error bars on open symbols are from ref. 10. The onset of charge order is identified by both to be $T_{CO}$ = 80 ± 10 K. *Lower right panels*: transport coefficients measured on a sample of Eu-LSCO with $p$ = 0.125: **f)** in-plane electrical resistivity ρ in a magnetic field $B$ = 0 (open symbols) and 15 T (closed symbols); **g)** Hall coefficient $R_H$ in 10 T; **h)** Nernst coefficient $v$ in 10 T. The onset of charge order is seen to coincide with anomalies in transport (marked by arrows): the minimum in ρ($T$), the drop in $R_H(T)$ and the sign change in $v(T)$, all at ~ 100 K. As for Nd-LSCO at $p$ = 0.20, this again argues for a Fermi-surface reconstruction caused by stripe order.

**Figure 2 | Doping evolution of the Nernst coefficient and $T_v$.**

Temperature dependence of the Nernst coefficient $v(T)$ for different dopings in Eu-LSCO [$p$ = 0.125 in green; $p$ = 0.16 in black] and Nd-LSCO [$p$ = 0.20 in red; $p$ = 0.24 in blue] at $B$ = 10 T. This shows the doping evolution of the two contributions to $v(T)$, respectively from superconducting fluctuations at low temperature and quasiparticles on a reconstructed Fermi surface at high temperature. The gradual convergence of the two peaks in $v(T)$ is a consequence of the fact that $T_v$ – the onset of the high-temperature peak (defined in Supplementary Fig. S2) – and $T_c$ – which controls the location of the low-temperature peak – come together as they approach $p^*$, the quantum critical point for the onset of stripe order[6] (see inset).

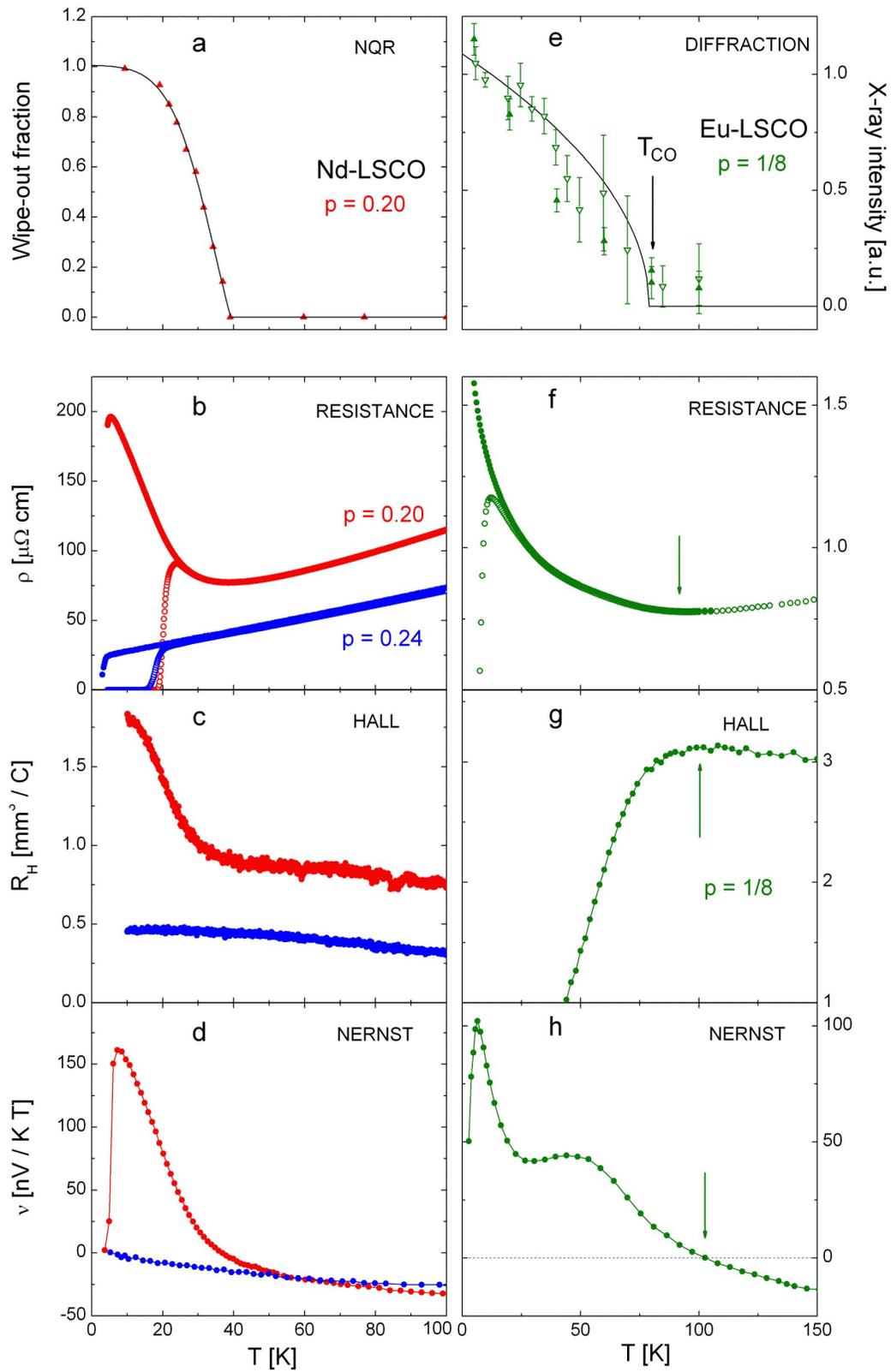

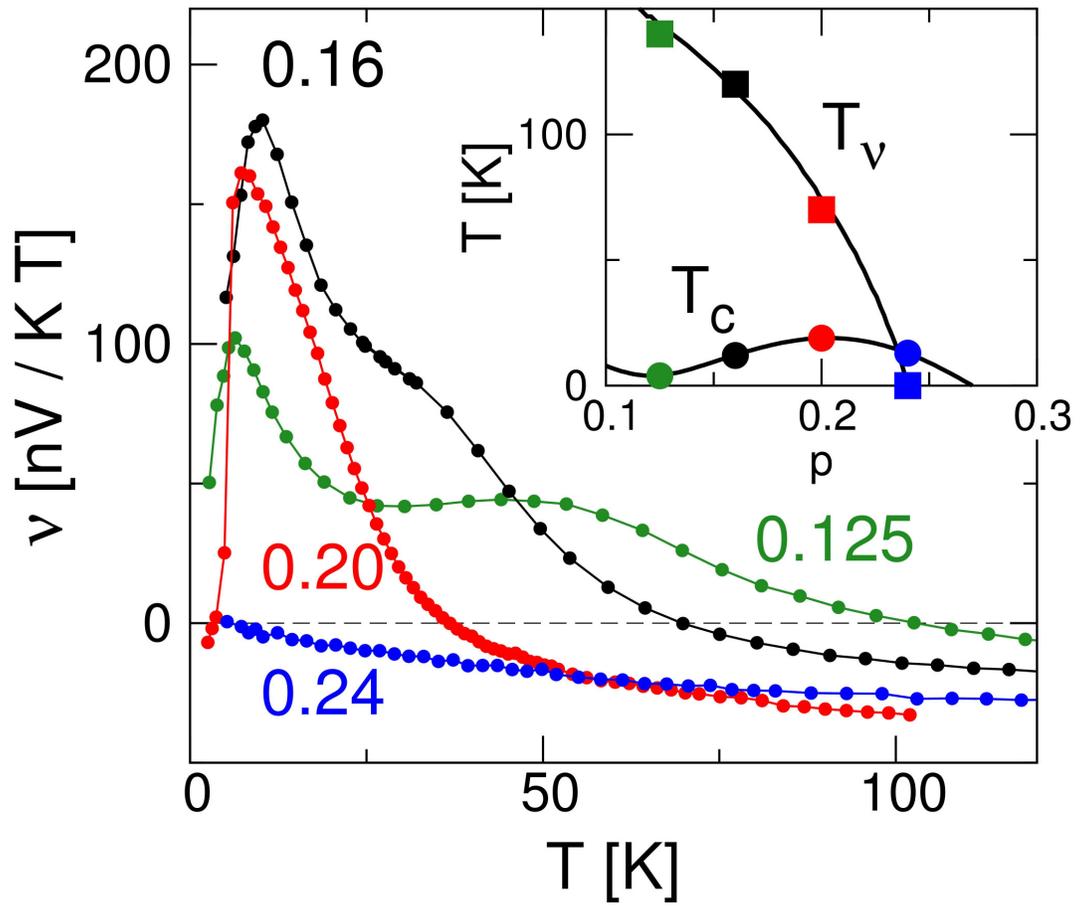

# Supplementary Information for

# "Enhancement of the Nernst effect by stripe order in a high-Tc superconductor"

Olivier Cyr-Choinière [1], R. Daou [1], Francis Laliberté [1], David LeBoeuf [1],

Nicolas Doiron-Leyraud [1], J. Chang [1], J.-Q. Yan [2,†], J.-G. Cheng [2], J.-S. Zhou [2],

J.B. Goodenough [2], S. Pyon [3], T. Takayama [3], H. Takagi [3,4], Y. Tanaka [5,3]

&  Louis Taillefer [1,6]

*1 Département de physique and RQMP, Université de Sherbrooke, Sherbrooke, Québec
J1K 2R1, Canada*

*2 Texas Materials Institute, University of Texas at Austin, Austin, Texas 78712, USA*

*3 Department of Advanced Materials, University of Tokyo, Kashiwa 277-8561, Japan*

*4. RIKEN (The Institute of Physical and Chemical Research), Wako, 351-0198, Japan*

*5. RIKEN SPring8  Center, Hyogo 679-5148, Japan*

*6. Canadian Institute for Advanced Research, Toronto, Ontario M5G 1Z8, Canada*

[†] Present address : Ames Laboratory, Ames, Iowa 50011 USA

## METHODS

**Nd-LSCO**. Single crystals of $La_{1.6-x}Nd_{0.4}Sr_xCuO_4$ (Nd-LSCO) were grown at the University of Texas using a travelling float zone technique. *ab*-plane single crystals were cut from boules with nominal Sr concentrations $x = 0.20$ and $x = 0.25$. The actual doping $p$ of each crystal was estimated from its $T_c$ and $\rho(250 K)$ values compared with published data, giving $p = 0.20 \pm 0.005$ and $0.24 \pm 0.005$, respectively.

**Eu-LSCO.** Single crystals of $La_{1.8-x}Eu_{0.2}Sr_xCuO_4$ (Eu-LSCO) were grown at the University of Tokyo using a travelling float zone technique, with Sr concentrations $x = 0.125$ and $x = 0.16$. The doping $p$ is taken to equal the Sr content $x$, to within $\pm 0.005$. The physical dimensions of the *ab*-plane samples cut out of the single-crystal boules were measured using an optical microscope and are shown in Table 1. The length $L$ is measured between the contacts used to measure the temperature difference or voltage drop along the current direction (*x*-axis).

**Table 1**

| Sample | Length, *L* [mm] | Width, *w* [mm] | Thickness, *t* [mm] |
|---|---|---|---|
| Eu-LSCO x=0.125 | 0.94 ± 0.10 | 0.28 ± 0.02 | 0.19 ± 0.02 |
| Eu-LSCO x=0.16 | 0.45 ± 0.10 | 0.43 ± 0.02 | 0.23 ± 0.02 |
| Nd-LSCO x=0.20 | 1.51 ± 0.05 | 0.50 ± 0.02 | 0.64 ± 0.02 |
| Nd-LSCO x=0.25 | 2.50 ± 0.05 | 0.51 ± 0.02 | 0.51 ± 0.02 |

**Superconducting transition temperature $T_c$**. The superconducting transition temperature $T_c$ of our Nd / Eu-LSCO samples was determined via resistivity measurements. In Table 2, we give $T_c$ values for two different criteria: 1) the temperature where the resistivity goes to zero; 2) the midpoint of the transition.

**Table 2**

| Sample | $T_c$ [K] ($\rho = 0$) | $T_c$ [K] (midpoint $\rho$) |
|---|---|---|
| Eu-LSCO x=0.125 | 5 ± 2 | 8 ± 4 |
| Eu-LSCO x=0.16 | 16 ± 3 | 24 ± 5 |
| Nd-LSCO x=0.20 | 20 ± 1 | 23 ± 3 |
| Nd-LSCO x=0.25 | 17 ± 1 | 20 ± 3 |

**Contacts**. Electrical contacts on the Eu / Nd-LSCO samples were made to the crystal surface using Epo-Tek H20E silver epoxy. This epoxy was cured for 5 min at 180 C then annealed at 500 C in flowing oxygen for 1 hr so that the silver diffused into the surface. This resulted in contact resistances of less than 0.1 Ω at room temperature. The longitudinal contacts were wrapped around all four sides of the sample. The current contacts covered the end faces. Nernst / Hall contacts were placed opposite each other in the middle of the samples, extending along the length of the $c$-axis, on the sides. The uncertainty in the quoted length $L$ of the sample (between longitudinal contacts) reflects the width of the voltage / temperature contacts along the $x$-axis.

**Measurement of the Nernst coefficient**. The Nernst signal was measured by applying a steady heat current through the sample (along the $x$-axis). The longitudinal thermal gradient was measured using two uncalibrated Cernox chip thermometers (Lakeshore), referenced to a further calibrated Cernox. The transverse electric field was measured using nanovolt preamplifiers and a nanovoltmeter. The temperature of the experiment was stabilized at each point to within ±10 mK. The temperature and voltage were measured with and without applied thermal gradient ($\Delta T$) for calibration. The magnetic field $B$, applied along the $c$-axis ($B \parallel z$), was then swept, with the heat on, from – 10 to + 10 T at 0.35 T / min, continuously taking data. The thermal gradient was monitored continuously and remained constant during the course of a sweep. The Nernst coefficient ($N$) was extracted from the part of the measured voltage antisymmetric with respect to magnetic field:

$$N = E_y / ( \partial T / \partial x ) = [ \Delta V_y(B) / \Delta T_x - \Delta V_y(-B) / \Delta T_x ] ( L / 2w ) ,$$

where $\Delta V$ is the difference in the voltage measured with and without thermal gradient. $L$ is the length (between contacts along the $x$-axis) and $w$ the width (along the $y$-axis) of the sample. This anti-symmetrization procedure removes any thermoelectric contribution from the sample or from the rest of the measurement circuit.

**Extraction of $T_\nu$.** We define $T_\nu$ as the point where $\nu / T$ deviates from linearity at high temperature; see Figures S2 and S5. This criterion is based on the fact that $\nu / T$ is linear in $T$ at all $T$ in Nd-LSCO at $p = 0.24 > p^*$, our reference sample where there is neither superconducting contribution to the Nernst signal nor any Fermi-surface reconstruction. This qualitative definition allows us to identify $T_\nu$ unambiguously to within +/- 10 K.

**Measurements of resistivity and Hall coefficient**. The resistivity $\rho(T) \equiv R_{xx} \, w \, t / L$ and Hall coefficient $R_H(T) \equiv R_{xy} \, t / B$ of each sample were measured using the standard six-terminal AC technique. A resistance bridge or a lock-in amplifier was used to measure the resistance. Field reversal was used to obtain the symmetric and anti-symmetric parts of the voltages, accounting for any misalignment of the contacts. Therefore, the longitudinal ($R_{xx}$) and transverse ($R_{xy}$) resistances were obtained as follows:

$$R_{xx} = ( R(B) + R(-B) ) / 2 \quad \text{and} \quad R_{xy} = (R(B) - R(-B)) / 2.$$

**Measurements of hard X-ray diffraction.** Hard X-ray diffraction measurements were performed with the BL19LXU beamline at RIKEN SPring-8. The photon energy was tuned to 24 keV. Q-scan profiles along the *h* direction revealed a broad superstructure reflection at $(4-2\varepsilon, 0, 0.5)$ with $2\varepsilon = 0.238(5)$ at low temperatures, indicative of stripe charge ordering. The peak was modelled with a Gaussian, assuming a linear background.

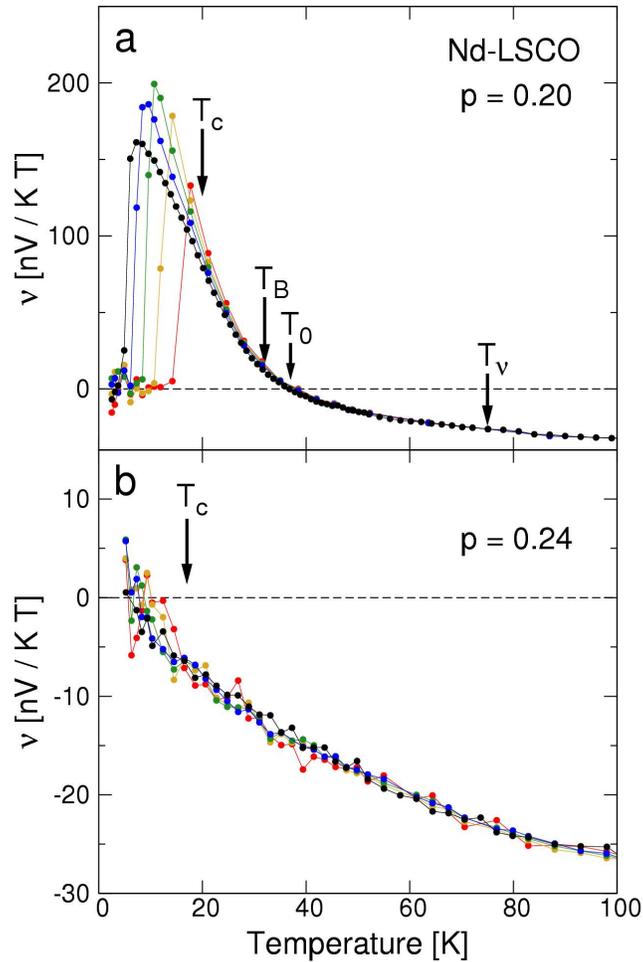

**Figure S1 | Effect of a magnetic field on the Nernst coefficient of Nd-LSCO.**

Nernst coefficient $v$ as a function of temperature for Nd-LSCO at $p = 0.20$ (upper panel) and $p = 0.24$ (lower panel), for different magnetic field strengths: $B = 2$ T (red), 4 T (yellow), 6 T (green), 8 T (blue), 10 T (black). $T_c$ is the zero-field superconducting transition (where $\rho = 0$). For $p = 0.20$, the onset of field dependence is labelled $T_B$. At higher temperature, $v / T$ becomes linear in temperature above $T_v$ (see Fig. S2). By contrast, for $p = 0.24$, the field dependence is within the noise of the measurement down to $T_c$ and both $T_B$ and $T_v$ are indistinguishable from zero.

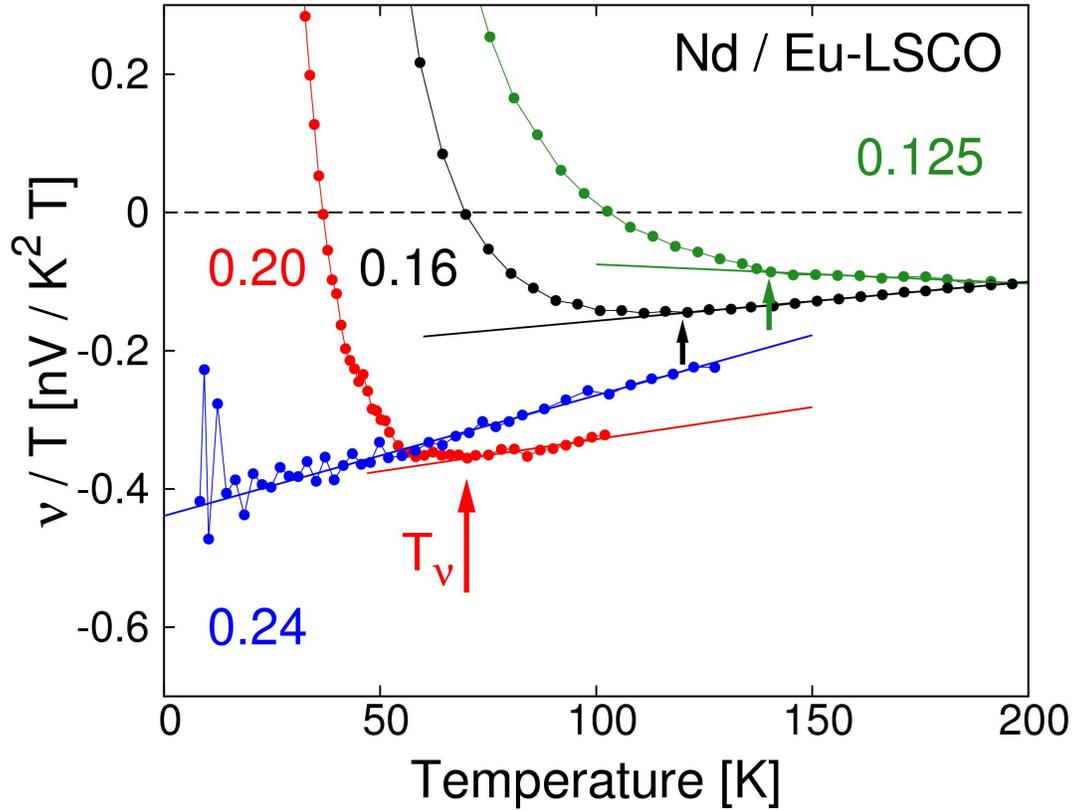

**Figure S2 | Onset of the positive upturn in the Nernst coefficient.**

Nernst coefficient *v* divided by temperature *T* for Eu-LSCO at *p* = 0.125 (green) and *p* = 0.16 (black), and for Nd-LSCO at *p* = 0.20 (red) and *p* = 0.24 (blue). All curves are taken in 10 T. The onset temperature $T_v$ (arrows) is defined as the deviation of *v* / *T* from a linear fit at high temperature. This yields $T_v$ = 140 ± 10, 120 ± 10, 70 ± 10 and 0 K, respectively.

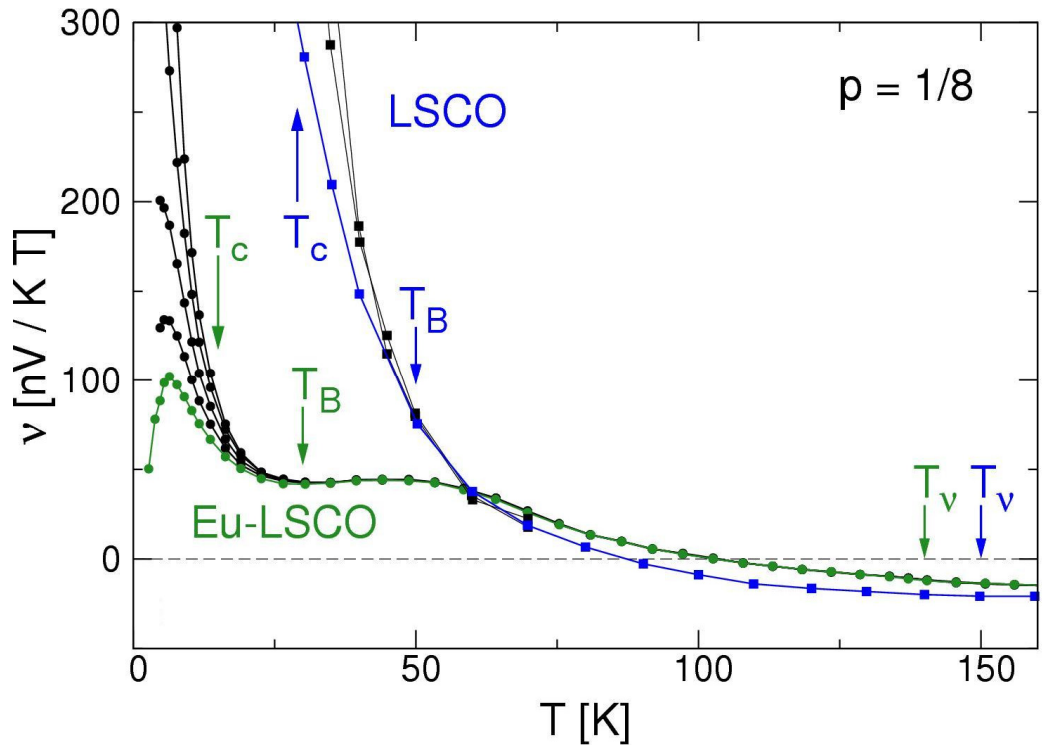

**Figure S3 | Comparison of Nernst coefficient in Eu-LSCO vs LSCO.**

Temperature dependence of the Nernst coefficient $v(T)$ for different magnetic fields in Eu-LSCO at $p = 0.125$ [circles] and LSCO at $p = 0.12$ [squares; from ref. 1]. Field strengths are 2, 4, 6, 8 and 10 T for Eu-LSCO (top to bottom), and 1, 6 and 14 T for LSCO (top to bottom). $T_v$ marks the onset of the positive rise at high temperature, as defined in Supplementary Figs. S2 and S5. $T_B$ marks the onset of a field dependence in $v(T)$, the expected signature of superconducting fluctuations. $T_c$ marks the onset of the superconducting transition in the zero-field resistivity. Note how $v(T)$ in Eu-LSCO exhibits two separate peaks, at 7 K and 45 K, which we attribute respectively to superconducting fluctuations (characterized by a strong field dependence) and quasiparticles (no field dependence), with respective onsets at $T_B$ and $T_v$. In LSCO at the same doping, the rise in $v(T)$ at high temperature is very similar, but the low-temperature field-dependent rise has moved up in temperature, with $T_B$ tracking $T_c$.

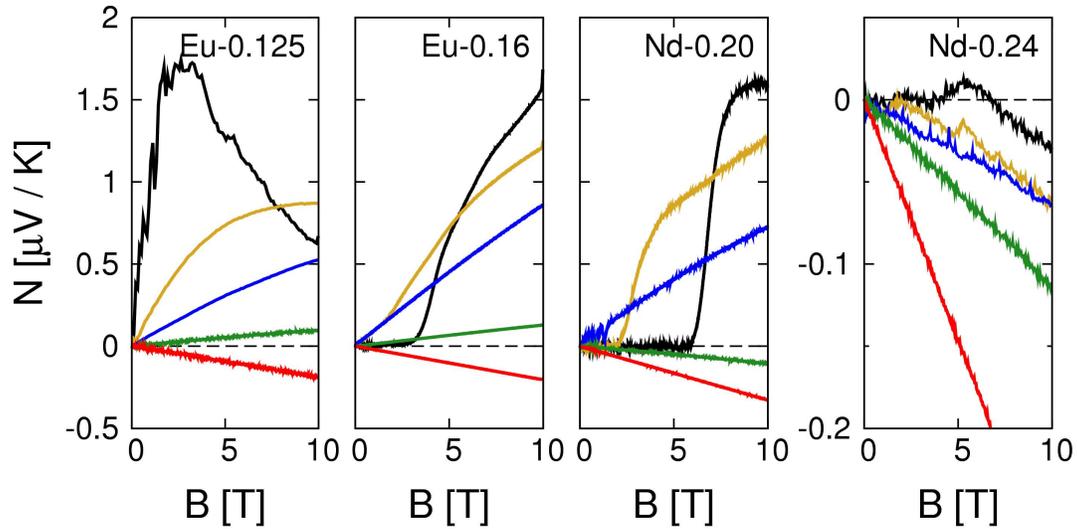

**Figure S4 | Magnetic field dependence of the Nernst signal.**

Field dependence of the Nernst signal in the four samples of Eu / Nd-LSCO measured in this study, at several temperatures: above $T_\nu$ [red]; between $T_\nu$ and $T_B$ (the onset of non linearity in $N$ vs $B$) [green]; between $T_B$ and the midpoint of the zero-field superconducting transition, $T_c$ [blue]; below $T_c$ [yellow and black]. The temperature of each curve is, respectively : 184, 83.6, 17.6, 9.7, 3.4 K ($p$ = 0.125); 196, 59.2, 32.0, 18.6, 7.2 K ($p$ = 0.16); 106, 45.4, 21.1, 14.2, 8.4 K ($p$ = 0.20); 132, 28.9, 16.5, 12.4, 8.3 K ($p$ = 0.24).

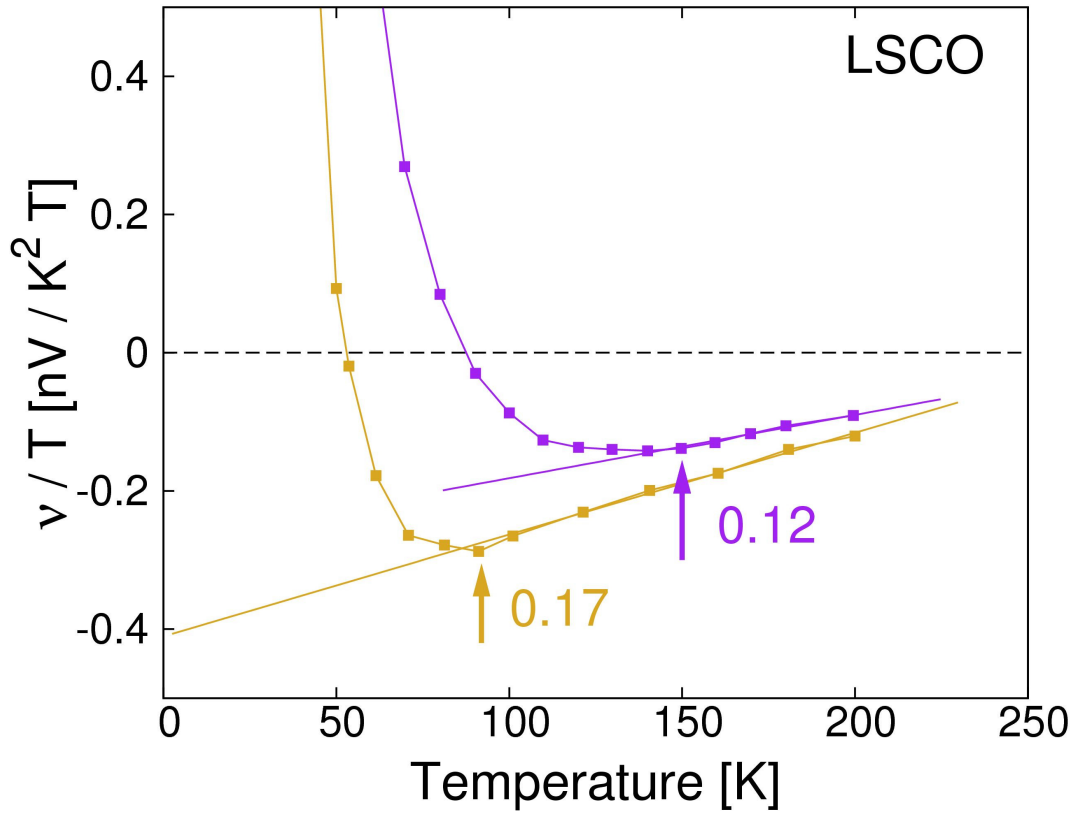

**Figure S5 | Onset of the positive upturn in the Nernst coefficient in LSCO.**

Nernst coefficient $v$ divided by temperature $T$ for LSCO at $p = 0.12$ (purple) and $p = 0.17$ (yellow) (from refs. 1, 2, and 3). Both curves show the zero field limit; there is no evidence of field dependence above 60 K. The onset temperature $T_v$ is defined as the deviation of $v / T$ from a linear fit at high temperature, giving $T_v = 150 \pm 10$ and $90 \pm 10$ K, respectively.

[1] Wang, Y., Li, P. & Ong, N.P. *Phys. Rev. B* **73,** 024510 (2006).

[2] Ong, N.P. *et al.*, *Ann. Phys. (Leipzig)* **13**, 9-14 (2004).

[3] Wang, Y. *et al.*, *Phys. Rev. Lett.* **88,** 257003 (2002).